\documentclass{JHEP3}
\usepackage{amsmath}
\usepackage{cite}
\usepackage{epsfig}
\usepackage{graphicx}
\def\be{\begin{equation}}
\def\ee{\end{equation}}
\def\baray{\begin{eqnarray}}
\def\earay{\end{eqnarray}}
\def\ba{\begin{eqnarray}}
\def\ea{\end{eqnarray}}

\title{DBI Global Strings}

\author{Saswat Sarangi \\
Institute of Strings, Cosmology and Astroparticle Physics \\
Department of Physics \\
Columbia University, New York, NY 10027, USA
}
\date{\today}
\abstract{In this note we present global string solutions which are a generalization of 
the usual field theory global vortices when the kinetic term is DBI.  
Such vortices can result from the spontaneous symmetry breaking in the potential felt by a 
D$3$-brane. In a previous paper (hep-th/$0706.0485$), the DBI instanton solution
was constructed which develops a "wrinkle" for stringy heights of the potential.
A similar effect is also seen for the DBI vortex solution. The wrinkle develops for
stringy heights of the potential. One recovers the usual field theory global string for 
substringy potentials. As an example of the symmetry breaking, 
we consider a mobile D$3$-brane on the warped deformed conifold. Symmetry breaking can occur
if the structure of the vacuum manifold of the potential for the D$3$-brane changes as it 
moves through the throat region.
}

\begin{document}

\section{Introduction}

String theory embeddings of inflationary scenarios typically involve  
D-branes which are either static or undergoing some motion in the transverse directions.
The world volume theory of the D-brane is described by the DBI action. 
Since the DBI action is a 
higher dimensional generalization of the Lorentz invariant relativistic action for 
a point particle, relativistic effects are to be expected
for the D-brane dynamics under appropriate relativistic conditions. Such examples of DBI 
relativistic effect are  studied in \cite{Brown:2007ce,Brown:2007vh,Silverstein:2003hf,
Chen:2004gc}. In \cite{Silverstein:2003hf}, the speed limit 
arising due to the relativistic nature of the DBI action is responsible for inflation.  
In \cite{Chen:2004gc} a variant of this scenario was proposed, 
again based on the DBI action. Further, the DBI effect has been shown to prevent the occurence of
slow roll eternal inflation in DBI inflation scenarios \cite{Chen:2006hs}. 
These inflationary scenarios employ mobile branes and, therefore,
when the D$3$-brane speed becomes relativistic see DBI effects. However, the
nonlinear nature of the DBI action can lead to new behavior in certain phenomena even when
the D-brane is static. In \cite{Brown:2007ce} the tunneling
of a D$3$-brane from a metastable to a true vacuum was investigated along the lines of the usual 
quantum field theory analysis given in \cite{Coleman:1977py}. For substringy barriers
the tunneling picture for the DBI action matches with the usual
QFT picture given in \cite{Coleman:1977py}. But once the barrier
between the metastable and the true vacuum assumes stringy heights,  
tunneling is much more enhanced than what one would have expected
based on the usual QFT intuition. This is surprising as the usual
QFT intuition has taught us to expect an exponential suppression
in the tunneling probability with an increase in the height of the barrier.
This counterintuitive result is easily understood, however, based on
a similar treatement of a point charged particle tunneling through an
electrostatic barrier \cite{Brown:2007vh}. Once the height of the barrier is large enough,
there is a significant Schwinger pair production of point particle-
antiparticle and this Schwinger effect can enhance tunneling. Similar effect
for the D$3$-brane enhances the tunneling rate. The D-brane, therefore, need not be 
moving at relativistic speeds to see interesting DBI effects. In this
paper we investigate global vortex solutions in the world volume of a D$3$-brane described by
the DBI action. Similar to the 
findings in \cite{Brown:2007ce}, these vortex solutions display a
departure from the usual field theory behavior once the height of the potential
of the complex scalar field becomes stringy. 

The DBI vortex solutions that we will construct should not be confused with the vortex solutions
of the tachyon field formed due to tachyon condensation after brane-antibrane annihilation 
(constructed, for example, in \cite{Jones:2002si} ) which correspond to codimension two branes.
The DBI vortices we construct are generalizations of the well known field theory global vortices
when the kinetic term is DBI. Indeed, for small gradients (which will correspond to small heights
of the potential) the DBI vortex resembles the usual field theory vortex. Further, these vortices
can form even when the separation between a brane and an antibrane is much greater than the critical
distance when tachyon condensation initiates. For the tachyon vortices to form, the brane-antibrane
separation should be string length.

\section{The Set-up}

The set up involves a $D3$-brane on a warped background. The DBI action for 
the $D3$-brane is given by \cite{Silverstein:2003hf}
\ba
\label{eom}
S = -\int d^4x \sqrt{-g}\left( f(\phi)^{-1}\sqrt{1+f(\phi)g^{\mu \nu}
\partial_\mu \phi \partial_\nu \phi} -f(\phi)^{-1} +V(\phi) \right)
\ea
where $f(\phi)$ is the warp factor. When the gradient terms are small,
one can expand the square root and keep the lowest order gradient term,  
the above action then reduces to the usual field theory action with the quadratic 
kinetic term. We are interested in vortex solutions of the DBI action. 

To study vortex solutions we must generalize the above action to the
case of a complex scalar field $\psi = \psi_1 + i \psi_2$. We consider
the action
\ba
\label{dbi}
S = -\int d^4x \sqrt{-g}\left( \sqrt{1+ g^{\mu \nu}
\partial_\mu \psi^* \partial_\nu \psi} - 1 +V(|\psi|) \right)
\ea
where $|\psi| =\sqrt{\psi\psi^*}$ is the absolute value. $V(\psi)$ has the
mexican hat shape, $V(|\psi|) = V_0 \left(\psi^* \psi -1 
\right)^2$. Note that $\psi$ does not have to be the radial coordinate
of Eq.(\ref{eom}). $\psi_1$ and $\psi_2$ could be the angular coordinates
at a fixed radial distance along the throat region of the warped deformed
conifold.

As an example, consider the warped compactification of type IIB string theory down
to four dimensions with metric ansatz
\baray
ds^2 = e^{2A(y)} g_{\mu \nu}dx^\mu dx^\nu + e^{-2A(y)} \tilde{g}_{mn}dy^m dy^n,
\earay
where $e^{A(y)}$ is the warp factor related to $f(\phi)$ in Eq.(\ref{eom}) by the relation $e^{4A(y)} = \alpha'^2/f(\phi)$. 
We consider compactifications
which are of the GKP type, which includes the Klebanov-Strassler warped throat arising as part of a compact
geometry. One begins with a conifold singularity with three-cycles $A$ and $B$ and complex structure modulus
$\epsilon^2 = \int_{A}\Omega$, $\Omega$ being the holomorphic three form. $\epsilon$ is stabilized by the fluxes
\baray
\int_A F_3 = M, \nonumber \\
\int_B H_3 = -K, 
\earay
at an exponentially small value $\epsilon \sim \exp(-\frac{\pi K}{g_s M})$.
The resulting geometry is well described by a KS throat, over which the warp
factor is strongly varying, attached to the rest of the compact space.
The unwarped metric of this region is that of a deformed conifold which comes to
a smooth end at the tip of the throat. Far from the tip, but still in the throat,
the unwarped metric is given by
\baray
\tilde{g}_{mn}dy^m dy^n \simeq d\rho^2 + \rho^2 ds^2_{T^{1,1}}, 
\earay 
$ds^2_{T^{1,1}}$ is the canonical metric on the five dimensional Einstein space $T^{1,1}$
, which is topologically $S^3 \times S^2$. At the tip of the throat, the $S^2$ shrinks to zero size and
the metric is well approximated by
\baray
\tilde{g}_{mn}dy^m dy^n \simeq \epsilon^{4/3}\left( d\tau^2 + \tau^2 d\Omega^2_2 + d\Omega_3^2\right), 
\earay
where $\rho^3 = \epsilon^2 cosh(\tau)$. At the tip $\tau = 0$.

In Eq.(\ref{dbi}), $\psi_1$ and $\psi_2$ could, for example, correspond to two angular directions
in the $S^3$ at a fixed value of $\rho$. Then $V(|\psi|)$ then corresponds to the potential for these 
angular locations of the D$3$-brane. In particular, the moduli space of a D$3$-brane/antibrane arising due
to nonperturbative Kahler moduli fixing effects has been constructed in \cite{DeWolfe:2007hd}. The potential felt 
by the D$3$-brane/antibrane along angular directions of the $S^3$ of the warped deformed conifold has been
explicitly constructed. Our choice of $\psi_1$ and $\psi_2$ corresponding to two angular coordinates on the $S^3$
is motivated by this construction. 

\section{Review of field theory global vortices} \label{qft}

When the $(\partial_\mu \psi)^2$ term is small, the action (Eq.(\ref{dbi}))
reduces to the usual field theory action for a complex scalar field
\ba
S = \int d^4x \sqrt{-g} \left(\frac{1}{2}g^{\mu \nu}\partial_\mu \psi^*
\partial_\nu \psi - V_0\left(\psi^*\psi-1 \right)^2 \right).
\ea
The equation of motion for $\phi$ is
\ba
\partial_\mu \partial^\mu \psi + 2V_0 (|\psi|^2-1)\psi =0.
\ea
One looks for a static axisymmetric solutions of the form
\ba
\psi(r) = e^{in\theta}\chi(r),
\ea
$\chi(r)$ satisfies the following second order nonlinear differential equation
\ba
\label{cs}
\frac{d^2\chi}{dr^2}+\frac{1}{r}\frac{d\chi}{dr}-\frac{n^2}{r^2}\chi - 2V_0
\chi(\chi^2-1)=0.
\ea
One requires that $\chi(r)$ satisfy the following boundary conditions
\ba
\chi(r) \to 1, r \to \infty, \nonumber \\
\chi(0) = 0.
\ea
The resulting solution is a global vortex with the asymptotics
\ba
\chi(r) \approx c_nr^n + ..., r \to 0, \nonumber \\
\chi(r) \approx 1 - \boldmath{O}(r^{-2}), r \to \infty.
\ea
The global vortex therefore corresponds to a static field configuration
that sits on the vacuum locus $\phi^*\phi = 1$ at infinity and at the
top of the potential at the core of the vortex (i.e. at $r  = 0$). 
Using numerical methods one can further determine the coefficients
$c_n$. For example, setting $V_0 = 0.5$, for $n=1$ winding, one gets
$c_1 \approx 0.58$. The energy per unit length of the global vortex
solution is logarithmically divergent.

\section{DBI global vortex solution}
We generalize the above analysis to the DBI action (Eq.(\ref{dbi})).
Using the ansatz for axisymmetric solutions $\psi = e^{in\theta}\chi(r)$,
the action (Eq.(\ref{dbi})) reduces to
\ba
S = \int dr r \left[ -\sqrt{1+ \left( \frac{d\chi}{dr}\right)^2 + 
\frac{n^2\chi^2}{r^2}} + 1 - V_0 \left(\chi^2 -1 \right)^2 \right].
\ea
The corresponding equation of motion for $\chi(r)$ is
\ba
\label{dbics}
\frac{d}{dr} \left(\gamma \frac{d\chi}{dr} \right) + \frac{\gamma}{r}
\frac{d\chi}{dr} - \gamma \frac{n^2\chi}{r^2} - 2V_0 \chi\left(\chi^2-
1 \right) = 0,
\ea
where the relativistic $\gamma$ factor is given by
\ba
\gamma = \frac{1}{\sqrt{1+ \left( \frac{d\chi}{dr}\right)^2 + 
\frac{n^2\chi^2}{r^2}}}.
\ea

Eq.(\ref{dbics}) is the DBI generalization of Eq.(\ref{cs}). We would like to 
construct vortex solutions to this equation, i.e. solutions satisfying the boundary
conditions $\chi(r=0)=0$, $\chi(r=\infty) =1$. However, as we will see, the DBI action will present 
us with difficulties while trying to impose one boundary condition at $r = 0$ and one at infinity.
Instead, we will be content with the (equivalent) boundary conditions at infinity $\chi(r=\infty)=1$,
$d\chi(r=\infty)/dr = 0$.  

Analytic solutions even for the relatively simple Eq.(\ref{cs}) are
not known, numerical methods have to be employed. Therefore, we shall simply try and find 
numerical solutions
to the much more complicated Eq.(\ref{dbics}). We construct numerical solutions
where $\chi(r \to \infty) = 1$ and $d\chi(r \to \infty)/dr =0$.

\subsection{Summary of the plots}

Before displaying the numerical results we summarize our findings.
We focus on winding number $1$, i.e. $n=1$.
We observe a phenomenon similar to that observed in the instanton
context in reference \cite{Brown:2007ce}. Starting with a small 
substringy value for $V_0$ (height of the mexican hat potential at 
$\chi =0$), we increase its value towards stringy
values. In the numerical plots, for $V_0 < 0.8$ we see the usual
global vortex solution, i.e. $\chi = 1$ at large  $r$ and $\chi = 0$
at the origin $r=0$. At the origin, just like in usual field 
theory, $d\chi/dr$ has some finite value.

 At $V_0 = 0.8$ (within the accuracy of these mathematica plots)
we observe $d\chi/ dr \to \infty$ close to (but not at) the origin. Further, the vortex
configuration fails to go all the way to the origin. $d\chi/dr$
becomes very large at around $r = 0.1$. This implies a ``turning
around'' is beginning to happen at this stage (similar to the
``wrinkling'' effect found in \cite{Brown:2007ce}).

 The plot for $V_0 = 0.9$ shows a ``turning around'' of the vortex
configuration, the field modulus $\chi$ has become double-valued near the origin. 
At infinity $\chi = 1$ as required for the vortex 
solution. As the radial distance $r$ decreases   
the field $\chi$ also drops towards $\chi=0$. However, because of
the DBI effect, before the field reaches the origin $r=0$,  the 
gradient $d\chi/dr$
becomes infinity. This implies a ``turning around'' of the field 
configuration. Now $r$, instead of decreasing anymore, increases
and $\chi$ keeps going towards $\chi = 0$. At $\chi = 0$, 
$d\chi/dr$ has some finite value and $r$ is finite. This 
configuration (taking into account the radial symmetry) looks
like a ``throat''. 

   Higher values of $V_0$ (plot for $V_0 = 1.5$) show the same behavior.
The usual field theory vortex has blown up into a throat solution. 

   For $V_0 = 2.0$ we see that at $\chi =0$ , $d\chi/dr \to -0$. 
So the solution at $r=\infty$ has $d\chi/dr = +0$ (and $\chi=1$).
As $r$ decreases, there is a turning around. After turning 
around the solution attains $d\chi/dr = -0$ when $\chi=0$.
This implies that the D-brane turns around , via the throat,
into an antibrane. Presumably such a configuration will be unstable.

\subsection{Plots}

\begin{figure}
\begin{center}
\includegraphics[width=6cm]{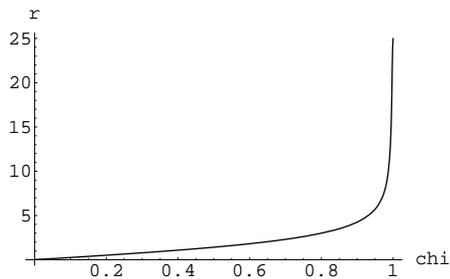}
\vspace{0.1in}
\caption{DBI global vortex solution with $V_0=0.1$. 
The vortex resembles a usual field theory global vortex.}
\label{fig1}
\end{center}
\end{figure}

\begin{figure}
\begin{center}
\includegraphics[width=6cm]{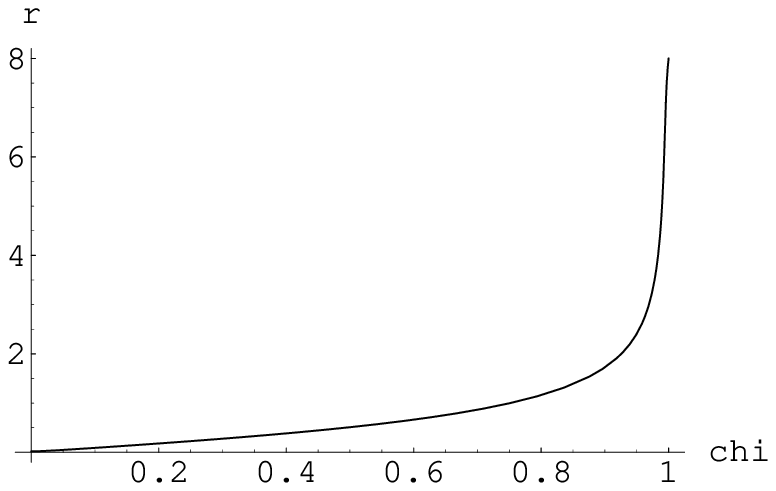}
\vspace{0.1in}
\caption{DBI global vortex solution with $V_0=0.5$. 
The vortex resembles a usual field theory global vortex.}
\label{fig2}
\end{center}
\end{figure}

\begin{figure}
\begin{center}
\includegraphics[width=6cm]{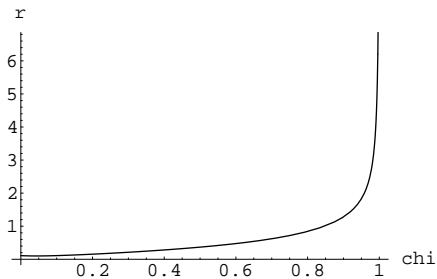}
\vspace{0.1in}
\caption{DBI global vortex solution with $V_0=0.8$. The gradient
$d\chi/dr \to \infty$ at $r=0.15$ and the field $\chi$ fails
to reach the origin $r=0$. At this value of $V_0$ the DBI
effect is siginificant.}
\label{fig3}
\end{center}
\end{figure}

\begin{figure}
\begin{center}
\includegraphics[width=6cm]{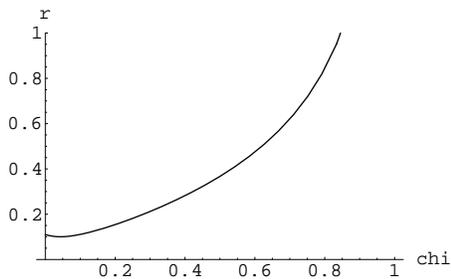}
\vspace{0.1in}
\caption{Once again DBI global vortex solution with $V_0=0.8$, but with 
the Y-axis range changed to show the ``turning around'' clearly. }
\label{fig4}
\end{center}
\end{figure}

\begin{figure}
\begin{center}
\includegraphics[width=6cm]{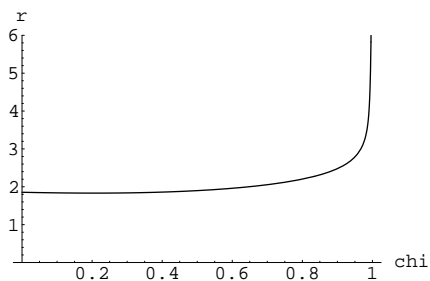}
\vspace{0.1in}
\caption{The DBI vortex solution with $V_0=0.9$. $d\chi/dr \to 0$
at large $r$, $d\chi/dr \to \infty$ at $r \approx 1.8$, then
$r$ increases again and $d\chi/dr$ becomes finite and negative
as $\chi \to 0$. }
\label{fig5}
\end{center}
\end{figure}

\begin{figure}
\begin{center}
\includegraphics[width=6cm]{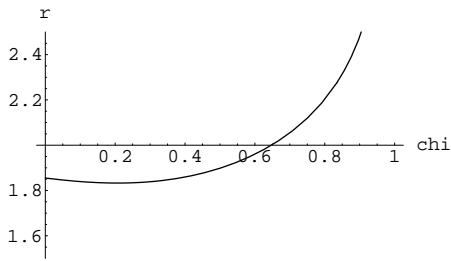}
\vspace{0.1in}
\caption{Once again DBI vortex solution with $V_0=0.9$. 
Y-axis range changed.}
\label{fig6}
\end{center}
\end{figure}

\begin{figure}
\begin{center}
\includegraphics[width=6cm]{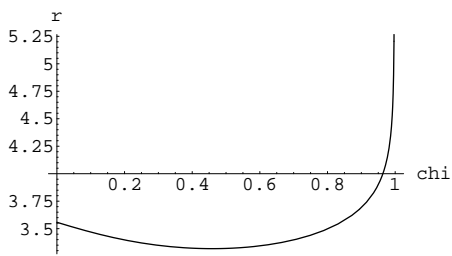}
\vspace{0.1in}
\caption{DBI global vortex solution with $V_0=1.5$.}
\label{fig7}
\end{center}
\end{figure}

\begin{figure}
\begin{center}
\includegraphics[width=6cm]{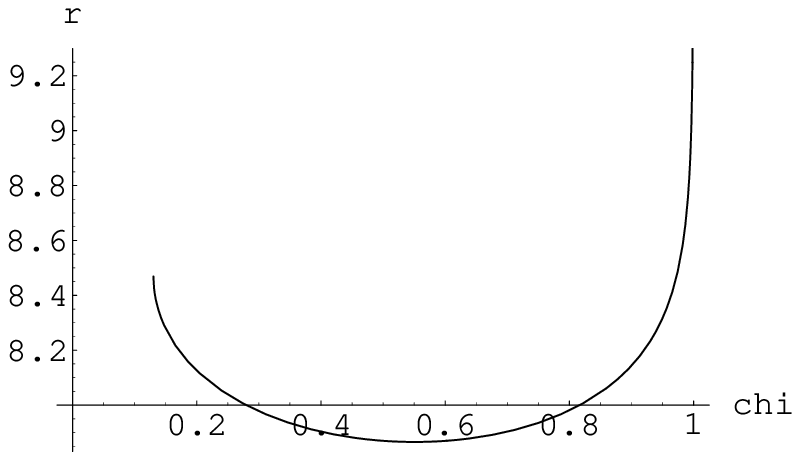}
\vspace{0.1in}
\caption{DBI global vortex solution with $V_0=2.0$. $d\chi/dr = -0$
as $\chi=0$ at the bottom of the throat. In other words, the D-brane
turns around and becomes an antibrane.}
\label{fig8}
\end{center}
\end{figure}

The mathematica plots of $r$ (the radial distance) versus $\chi$ are shown in
Figs(\ref{fig1} - \ref{fig8}).


\section{The D$3$-brane moduli space and formation of vortices} \label{moduli}

In \cite{DeWolfe:2007hd}, D$3$-brane and antibrane vacua on the 
warped deformed conifold were constructed. D$3$-branes do not feel
any force in no-scale flux compactifications of type II-B string theory.
However, the non-perturbative effects required to stabilize the Kahler
moduli destroy the no-scale feature and generate a potential for the
D$3$-brane. The shape of the potential depends on the compactification
geometry and on the embedding of the moduli-stabilizing branes. 
In \cite{DeWolfe:2007hd}, various supersymmetric embeddings of the 
D$7$-brane have been considered in the context of the warped deformed
conifold. These embeddings of the D$7$-brane generate a non-perturbative
superpotential which in turn generates the D$3$-brane vacua. The vacua
have real dimensions zero, one and two. The zero and one dimensional vacua
are generic in a compact Calabi-Yau. Furthermore, the D$3$-branes and
D$3$-antibranes share the same vacua at the tip of the warped deformed
conifold (which is important for the success ending of brane-antibrane inflation).
The authors give explicit moduli space of the vacua by considering the 
warped deformed conifold. At a fixed radial coordinate the deformed conifold
has a $S^2 \times S^3$. The $S^2$ shrinks to a zero size at the tip. 
The zero, one and two dimensional vacua correspond to a point, $S^1$ and
$T^2$ on the $S^3$. 

\begin{figure}
\begin{center}
\includegraphics[width=8cm]{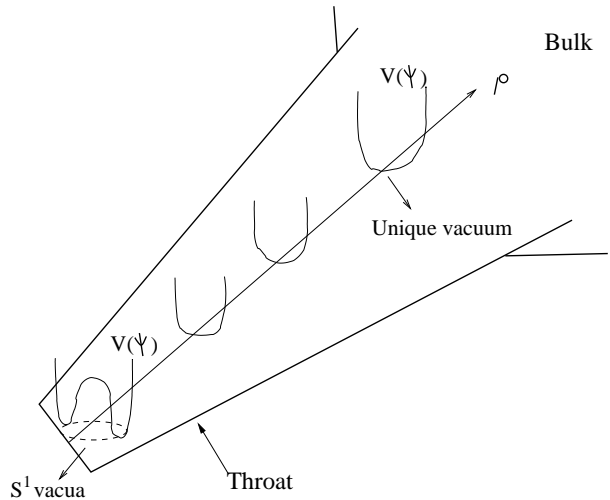}
\vspace{0.1in}
\caption{The throat region of a warped deformed conifold. $\psi_1$ and $\psi_2$ are two angular coordinates on the $S^3$
and together they constitute the complex scalar field $\psi$. The D$3$-brane feels the potential $V(\psi)= m(\rho)^2|\psi|^2 
+ \lambda |\psi|^4 $.  The $\rho$ coordinate represents the radial direction along the throat, and is assumed to be a flat direction.
$m(\rho)^2> 0$ away from the tip and $m(\rho^2) < 0$ at the tip. Away from the tip, $V(\psi)$ has a unique vacuum. At the tip, the degenerate 
vacua of $V(\psi)$ constitute a $S1$. The spontaneous symmetry breaking occurs as the D$3$-brane moves towards and 
eventually sits at the tip.}
\label{conifold}
\end{center}
\end{figure}

 The DBI vortex solutions constructed from the action in Eq.(\ref{dbi}), 
require the vacua to constitute a periodic one-dimensional loci (i.e. a $S^1$). 
This potential is modeled by the mexican hat potential of Eq.(\ref{dbi}). 

 The particular symmetry breaking mechanism that we have in mind is
 depicted in Fig.(\ref{conifold}). As an example of the symmetry breaking, 
we consider a mobile D$3$-brane on the warped deformed conifold. Spontaneous symmetry breaking can occur
if the structure of the vacuum manifold of the potential experienced by the D$3$-brane changes as it 
moves through the throat region. We consider two angular directions $\psi_1$ and
 $\psi_2$ transverse to the D$3$-brane world volume ($\psi^1$ and $\psi^2$ are some 
 coordinates on the $S^3$, for example). Together they constitute
 a complex scalar field $\psi = \psi_1 + i \psi_2$. The D$3$-brane potential $V(\psi)$  
is a function of these two transverse directions. Apart from these two directions $\psi_1$ 
and $\psi_2$, the usual radial direction $\rho$ along the warped deformed conifold is also present. 
The other transverse directions are not involved in the dynamics and are suppressed. We assume
that away from the tip (nonzero radial coordinate $\rho$), the vacuum is zero dimensional (i.e. there is a unique
minimum) and at the tip ($\rho=0$) the vacua constitute an $S^1$. I.e., the potential
has the form $V(\psi) = V_0 + m(\rho)^2 |\psi|^2 + \lambda |\psi|^4$, with $m(\rho)^2 > 0$ away
from the tip, and $m(\rho=0)^2 < 0$ at the tip. The spontaneous symmetry
breaking occurs as the D$3$-brane moves down the throat region towards the tip.
Away from the tip the D$3$-brane sees a unique vacuum and at the tip it can
settle down to any point of the $S^1$. From the D$3$-brane world volume perspective,
the dynamics of the complex field $\psi$ is described by the four dimensional
effective action of Eq.(\ref{dbi}). If we replace the DBI kinetic term by the
usual quadratic kinetic term $1/2\partial_\mu \psi \partial^\mu \psi^*$, then we are
just looking at the usual second order phase transition of a complex $\lambda |\psi|^4$ theory
that can form vortices. The radial coordinate $\rho$ serves as the symmetry breaking parameter that
controls the shape of the potential.

Points that are far away in the 
D$3$-brane world volume will be uncorrelated regarding their choice of the 
particular point on the $S^1$ to which they eventually settle down (the 
correlation length being the inverse mass scale $1/m$). Hence once the brane
settles down at the tip, its world volume will consist of various domains
with different values of the complex field $\psi$ (all belonging to the
$S^1$ moduli space). As a result vortices will form.
 
 Although we have been referring to the vortex formation due to the D$3$-brane dynamics, the
 same applies to D$3$-antibrane as well. As discussed in \cite{DeWolfe:2007hd}, one can construct
 the moduli space for the antibrane on the warped deformed conifold and just like the D$3$-brane
 case, the corresponding vacua have real dimensions zero, one and two. At the tip of the warped
 deformed conifold, the D$3$-brane and the D$3$-antibrane share the same vacua. This is important for ending
 brane inflation. In brane inflation scenarios, the antibrane sits at the tip of the warped deformed
 conifold while the D$3$ brane falls towards the tip due to the attraction it feels from the antibrane.
 Can we apply the above spontaneous symmetry breaking mechanism to brane-antibrane inflation and produce
 DBI strings at the end of inflation? The antibrane will first feel the attraction from the tip region and
 migrate towards the tip. Assuming the presence of $S^1$ vacua at the tip, the antibrane will not prefer any one
 point of this $S^1$ over any other and the vortices will form. Next the D$3$-brane will start falling towards
 the tip due to the brane-antibrane attraction. During this period inflation will take place, the relative separation 
 of the brane-antibrane system will play the role of the inflaton. This inflationary period will dilute away the
 DBI vortices. Further, the domains with different values of the complex $\phi$ field (all belonging to the $S^1$) 
 that formed on the world volume of the D$3$-antibrane will become exponentially large during the inflationary epoch.
 Hence towards the end of inflation, the D$3$-brane will effectively fall towards one point on the $S^1$ vacua
 inside each exponentially large region (which will contain the visible universe). Hence, within each exponentially
 large region there is no spontaneous symmetry breaking and no formation of vortices. In short, the global vortices
 will form when the D$3$-antibrane settles at the tip region. However, these vortices will get diluted during
 inflationary epoch. There will be no vortex formation when inflation ends and the D$3$-brane settles at the tip
 region as within eponentially large regions (the domains on D$3$-antibrane inflated to exponentially large
 sizes during inflation) the D$3$-antibrane sits at a unique point on the $S^1$ vacua. At this point
 the usual cosmic string production still happens due to brane-antibrane annihilation \cite{Sarangi:2002yt, Polchinski:2004ia}. 

\section{Comparison with k-defects}

The distinguishing features of DBI global vortex solution are due to
the square root kinetic term in the action. A general analysis of the defects 
arising from a class of actions with a nonlinear kinetic term has been done 
by Babichev in Refs \cite{Babichev:2006cy} and \cite{Babichev:2007tn} - 
these defects are called the {\it k-defects}. In these models, the kinetic 
term of the scalar field has the general form $M^4 K(X/M^4)$, where $X = 1/2 
(\partial_{\mu} \phi)^2$ and $M$ is some mass scale (called the {\it kinetic 
mass} scale) other than the mass scale $\eta$ that arises in the potential 
term $V(\phi) = (\phi^2 - \eta^2)^2$.  The size and energy density of the
k-defects is set by a combination of $M$ and $\eta$ which depends on the 
exact functional form of the kinetic term $K(X/M^4)$. For example 
(see \cite{Babichev:2006cy}), for a simple power law $K = -(X/M^4)^\alpha$, 
the size of the defect is given by the scale $M^{-1} (\eta/M)^{1-2/\alpha}$.  

This analysis of k-defects can be translated to our DBI case easily. The 
kinetic mass scale is simply the warp factor $f(\phi)$. The potential
energy $V(\phi)$ should be measured in units of this warp factor. Hence,
the effect of the DBI action becomes important only when $f(\phi)V(\phi)$
is order unity. Indeed, in the case of k-defects $M$ supplies an additional
mass scale which can combine with $\eta$ (the GUT scale, say) to give an
effectively low tension for the k-defects. This is similar to the 
observation that in a warped background the effective tension of a cosmic
string can be low due to the warp factor $f(\phi)$.

The appearance of the wrinkle for the DBI vortex, however, is a feature
that is essentially due to the relativistic nature of the square root 
kinetic term. In the somewhat different context of tunneling of D-branes
\cite{Brown:2007ce}, the appearance of the wrinkle is explained as due to the 
Schwinger effect. This effect, therefore, is not seen in the general
k-defect analysis of \cite{Babichev:2006cy}.

\section{Discussion}

We have considered the DBI action for a complex field $\psi$ with a 
potential term $V(\psi)$ and constructed
the vortex solutions which are generalizations of the usual global vortex 
of the complex scalar field theory. For substringy heights of the potential the vortex resembles
the usual field theory global vortex. For stringy potentials the vortex develops
a wrinkle analogous to the instanton wrinkle found in \cite{Brown:2007ce} \footnote{Vortex solutions for the DBI action have 
been studied by various authors including \cite{Callan:1997kz, Gibbons:1997xz, Hashimoto:1997px,Hashimoto:2003pu}. 
These vortex solutions (BIons) can develop a throat and have double valued $\phi$.}. 
We have neglected the world volume gauge fields. Including the world volume
gauge fields might lead to the DBI analogues of field theory local strings.
These vortices should not be confused with the vortex solutions on a tachyon profile
that are produced upon tachyon condensation initiated by the brane-antibrane instability.
Unlike tachyon vortices, the DBI vortices can form even when the brane-antibrane separation
is large in string length units.

 A similar effect was seen in the context of DBI instanton 
in \cite{Brown:2007ce} where the instanton
interpolating between the metastable and the true vacuum of the
potential $V(\phi)$ was studied in the thin wall limit. When the height of
the barrier $V_0 < 1$ one gets the usual QFT instanton. However
when $V_0 > 1$ the instanton develops a wrinkle due to the 
double valuedness (and turning around) of the instanton configuration.
For the vortex solution the wrinkle appears at $V_0 \approx 0.8$. This 
difference in the critical $V_0$ is due to the friction term present in
the vortex equation of motion (Eq.(\ref{dbi})). The wrinkled instanton
in \cite{Brown:2007ce} was constructed in the thin-wall limit where 
the friction term is set to zero. 

 The appearance of the wrinkle can be understood by following the analogy with
 the DBI instanton of \cite{Brown:2007ce}. Assuming
 that the vortex configuration will always have a large enough value of the
 radial coordinate $r$ (i.e. apriori assuming the formation of a wrinkle), 
 the second and the third terms in Eq.(\ref{dbics})
 (i.e. $ \frac{\gamma}{r} \frac{d\chi}{dr}$ and $\gamma \frac{n^2\chi}{r^2} $)
 can be neglected. The equation of motion then resembles the thin wall equation
 of motion for the DBI instanton in \cite{Brown:2007ce} and following the reasoning
 for the instanton, a wrinkle must develop. This is, of course, a heuristic way to
 see the appearance of a wrinkle in the vortex, which will fail to give the exact
 value of the radial coordinate $r$ or height of the potential $V_0$ where the wrinkle appears.
 
 The asymptotics of the DBI vortex at radial infinity will remain the same as that for
 the usual field theory vortex of Sec.(\ref{qft}). This is because at radial infinity the gradient
 term vanishes and the DBI kinetic term does not play any role. The DBI effect only occurs near
 the origin (i.e. near the core of the vortex) where the gradient term is big and sources the
 DBI term. Since the DBI global vortex asymptotics at radial infinity are the same as the usual field theory
 global vortex asymptotics at infinity, there will still be a logarithmic divergence in the
 total energy of the field configuration. However, as explained before, inspite of the 
 double-valuedness of the DBI vortex field configuration, the Hamiltonian density will remain
 finite everywhere.
 
 In Sec. (\ref{moduli}) we considered the possibility of the cosmological formation
 of these vortices in a brane inflation scenario. Brane inflation consists of two steps.
 First, before the inflation can begin, a D$3$-antibrane migrates down the throat and
 settles at the tip. In the presence of an appropriate moduli space, vortices can form
 on the antibrane world volume when it settles at the tip. However, in the next stage 
 a D$3$-brane is attracted to the antibrane and this leads to inflation. The inflationary
 stage washes away the vortices. Further, the different domains on the antibrane which 
 have the $\psi$ field at different points of the $S^1$ become exponentially large during inflation.
 Consequently, the D$3$-brane effectively falls towards a single point on the $S^1$. No DBI
 vortices form at the end of inflation. 
 
 Our approach in this note has been phenomenological. Motivated by the results for the D$3$-brane
 moduli space on a warped deformed conifold in \cite{DeWolfe:2007hd}, we consider a potential
 $V(\psi)$ along the $S^3$ angular coordinates for the D$3$-brane and examine the global vortex solution. 
 Whether or not a wrinkle will form depends on the height of the potential. If $V(\psi)$ never attains 
 stringy heights, then there is no possibility of a wrinkle. At this point we simply note that the simplest
 DBI inflation scenario \cite{Silverstein:2003hf} considers a DBI action of the form Eq.(\ref{eom}) (with 
 gravity added). The inflaton potential $V(\phi)$ is generated by radiative or bulk effects whose 
  value must be large when measured in terms of the warp factor/local string units $f(\phi)$. In particular, power law inflation   
 occurs at late times (i.e. when $t \to \infty$) as the mobile D$3$-brane is nearing the tip. The warp factor
 and the potential have late time dependence $f(\phi) \to t^4$ and $V(\phi) \to 1/t^2$ which leads to
 $f(\phi)V(\phi) \to t^2$ in the late time DBI regime. In such a setting it appears that one can generate 
 large potentials needed for interesting DBI effects. For the appearance of a wrinkle we require a potential
 whose value is order one in local string units. Another issue while considering the appearance of the wrinkle
 is the trustability of the DBI description. The DBI description is valid whenever the extrinsic curvature of the 
 solution $\psi(r)$ is low. In the absence of any warp factor this extrinsic
 curvature is given by \cite{Brown:2007ce, Sarangi:2007jb}
 \baray
 K(\psi) = \frac{1}{\sqrt{\alpha'}} \frac{\partial V(\psi)}{\partial \psi},
 \earay
 i.e. as long as the slope of the potential is small in string units, the DBI action has small higher derivative 
 corrections.

 It would be interesting to see if there is some version
 of brane inflation on a warped deformed conifold which can lead to the formation of these
 DBI vortices. Further, it would be interesting to find the local string version of these DBI global 
 vortices as local strings can be realistic cosmic string candidates. If such local strings still develop
 a wrinkle due to the DBI effect, this wrinkle would be the result of the stringy DBI action. 
 Such cosmic strings would, in principle, differ from their field theory cousins and perhaps lead to 
 novel phenomenological consequences.
   
   Further one could construct DBI defects, that are extensions of the DBI vortices we have studied, on the
   world volume of multiple D$3$-branes. The theory would then be a non-abelian one. One could then look
   , for example, for the DBI extension of the t' Hooft-Polyakov monopole. It is tempting to speculate that
   just as what we saw for the DBI vortex solution, there will be no DBI effects present at radial infinity 
   of the monopole solution. This is because the field gradient vanishes at large distances from the core of the monopole. The DBI
   effect would occur for large field gradients near the core of the monopole and lead to the formation of
   a wrinkle near the core for potentials with stringy heights. 
 
\acknowledgments

The author would like to thank Jacques Distler, Dan Kabat, Ben Shlaer, Henry Tye,
Bret Underwood and especially Gary Shiu for helpful discussions, and the theory
group at University of Texas, Austin, for hospitality while the work was in progress. Special thanks are due to Mark Wyman for key discussions that led to
this work and to Eugeny Babichev for bringing to the author's attention 
work on k-defects. This work was supported by the DOE contract DE-FG02-92ER40699.

\appendix

\section{Appendix: Numerical calculation and the boundary conditions}

The plots for the vortex configuration ($\chi$ versus $r$) were obtained by numerically solving for
the equation of motion for the scalar field modulus $\chi$ given in Eq.(\ref{dbics}). The boundary
conditions are given by $\chi(r=\infty) = 1$, and $\chi(r=0) =1$. While this is straightforward to
do for small values of $V_0$, once $V_0$ exceeds $~ 0.8$, the field configuration becomes double
valued and this numerical method for obtaining the solution fails. To get around this problem
appearing due to the double-valuedness of $\chi$, we rewrite the equation of motion with 
the radial coordinate $r$ as a function of the field modulus $\chi$, i.e. $r(\chi)$. One can do this by starting
with the action in Eq.(\ref{dbi}). With the ansatz $\psi(r,\theta) = e^{i n \theta}\chi(r)$, we get
\baray
S = -2\pi \int dr r \left[ \sqrt{1+\left(\frac{d\chi(r)}{dr}\right)^2 + \frac{n^2 \chi(r)^2}{r^2}} -1 + V(\chi)\right].
\earay
The equation of motion obtained from this action can be solved to obtain $\chi$ as a function of $r$. However,
we can rewrite the action as follows to have $r$ as a function of $\chi$
\baray
S = -2\pi \int d\chi r(\chi) \left[ \sqrt{1 + \left(\frac{dr(\chi)}{d\chi}\right)^2\left[ 1 + \frac{n^2\chi^2}{r(\chi)^2}\right]}
+ \left( V(\chi)- 1\right)\frac{dr(\chi)}{d\chi} \right] ,
\earay
whose variation leads to the equation of motion for $r$ as a function of $\chi$
\baray
\label{dual}
r(\chi) \frac{d}{d\chi}\left[ \frac{\left(1 + \frac{n^2\chi^2}{r(\chi)^2}\right)}{\sqrt{1+\left(1 + \frac{n^2\chi^2}{r(\chi)^2}\right)
\left(\frac{dr(\chi)}{d\chi}\right)^2}}\frac{dr(\chi)}{d\chi} \right] + \frac{\left( \frac{n^2\chi^2}{r(\chi)^2} \left(\frac{dr(\chi)}{d\chi}\right)^2-1 \right)}{\sqrt{1+\left(1 + \frac{n^2\chi^2}{r(\chi)^2}\right)
\left(\frac{dr(\chi)}{d\chi}\right)^2}} \nonumber \\
+ r(\chi)\frac{d V(\chi)}{d\chi} = 0.
\earay
This equation can be solved numerically. The plots shown in Figures (\ref{fig1}) to (\ref{fig8})
were obtained by numerically solving this equation subject to the boundary conditions $r(\chi=1) = \infty$,
and $dr(\chi = 1)/d\chi = \infty$ (where $\infty$ was replaced by sufficiently large numbers in the mathematica
numerical calculation).

A few words about the boundary conditions. In the usual coordinates ($\chi$ as a function of $r$) the boundary
conditions are taken to be $\chi(r=\infty) = 1$ and $\chi(r=0) = 0$. This translates to $r(\chi=1)= \infty$
and $r(\chi=0) = 0$. However, this leads to a problem for the DBI case. For large values of the potential (roughly $V_0 \geq 0.8$)
$\chi$ becomes double valued as a function of $r$ (although $r(\chi)$ still remains single valued as a function
 of $\chi$) and the field configuration never reaches $r = 0$. Instead, the configuration terminates at some
 finite value of $r$ at $\chi = 0$. The boundary conditions which are appropriate to describe this behavior are given
 by $r(\chi = 1) = \infty$ and $dr(\chi = 1)/d\chi = \infty$. These two boundary conditions give the usual QFT 
 vortex solutions for $V_0 << 0.8$ where the field configuration sits at the vacuum at radial infinity and sits
 at the top of the potential barrier at radial origin. However for large barriers ($V_0 \geq 0.8$) the "wrinkle"
 develops and the field configuration starting at the vacuum at radial infinity never quite makes it to
 the radial origin. 
 
 Qualitatively one can see why this happens. Let us go back to the usual description of the field where we consider
 $\chi$ as a function of $r$. From the equation of motion Eq.(\ref{dbics}), regarding $r$ as a time coordinate,
 we see that $\chi$ is moving in the inverted mexican hat potential. Our boundary conditions ($r(\chi = 1) = 
 \infty$ and $dr(\chi = 1)/d\chi = \infty$) correspond to field sitting at a maximum of the inverted mexican hat
 at $\chi = 1$ at $r = \infty$ and then rolling down towards the minimum of the inverted potential at $\chi = 0$.
 As the field rolls down and approaches the origin at $\chi = 0$, the relativistic factor $\gamma$ decreases. 
 At some point, the speed becomes infinite and $d\chi/dr$ becomes double valued. The energy density of the configuration
 is still finite. This is because we are really looking at a static solution and our $\gamma$ is a static analogue
 of the Lorentzian $\gamma$ factor. For brane dynamics (as discussed in \cite{Silverstein:2003hf}), the Lorentzian
 $\gamma$ factor diverges as the brane speed approaches the speed of light and, consequently, the Hamiltonian also
 diverges. However in our case, when the "speed" $d\chi/dr$ diverges, $\gamma$ vanishes.
 Consequently there is no divergence in the Hamiltonian density.

  One could ask what would have happened if we had tried to solve numerically the equation of motion Eq.(\ref{dbics})
 by specifying the two boundary conditions at the origin (i.e. at $r = 0$ considering $\chi(r)$ as a function
 of $r$). One can certainly specify the boundary conditions in this way : start with $\chi(r=0)=0$ and $d\chi(r=0)/dr
 = c$ where $c$ is a positive number. One can slowly increase $c$ from zero. Note that this corresponds
 to launching the field modulus $\chi$ at speed $c$ starting from the bottom of the inverted potential at $\chi=0$ towards
 the local maximum at $\chi=1$. We would like $\chi$ to reach this maximum with zero speed at $r=\infty$ and sit at this 
 unstable maximum. For $V_0 < 0.8$, increasing $c$ from zero upwards, one eventually finds the value of $c$ which 
 leads to this field configuration which corresponds to the vortex solution. However, once $V_0 = 0.8$ (approximately)
 no finite value of $c$ is large enough to take $\chi$ to the maximum of the inverted potential. This happens due
 to the relativistic effect which manifests as the friction term in the equation of motion. The infinite value of $c$ for
 $V_0 = 0.8$ implies that the solution $\chi(r)$ is about to become double valued as a function of $r$.
 For $V_0 > 0.8$ no solutions
 exist satisfy the boundary conditions $\chi(r=0) = 0$ and $\chi(r=\infty)=1$. Solutions exist where the field $\chi$ sits
 at the vacuum configuration (i.e. $\chi = 1$) at radial infinity with zero energy density at infinity (i.e. $d\chi(r=\infty)/dr =0$).
 But these solutions never interpolate all the way to $r = 0$. Furthermore these solutions are double valued in $\chi$ and
 we have to switch to Eq.(\ref{dual}) to be able to construct these numerically.


\end{document}